\documentclass[12pt]{article}
\usepackage[utf8]{inputenc}
\usepackage{graphicx,amssymb,bbm,amsmath,mathrsfs,hyperref,slashed}
\newcommand{\ket}[1]{\ensuremath{\left|#1\right\rangle}}

\title{The problem with `The Problem of Time'}
\author{\large K.L.H. Bryan${}^{(1)}$,  A.J.M. Medved${}^{(1,2)}$
\\
\vspace{-.5in} \hspace{-1.5in} \vbox{
 \begin{flushleft}
$^{\textrm{\normalsize (1)\ Department of Physics \& Electronics, Rhodes University,
  Grahamstown 6140, South Africa}}$
$^{\textrm{\normalsize (2)\ National Institute for Theoretical Physics (NITheP), Western Cape 7602,
South Africa}}$
\\ \small \hspace{1.07in}
  g08b1231@ru.ac.za,\  j.medved@ru.ac.za
\end{flushleft}
}}
\begin{document}
\date{}
\maketitle
 \begin{abstract}
    We investigate three aspects of  the supposed problem of time: The
    disagreement between  the treatments of time in general relativity and
    quantum theory,  the problem of recovering time from within an isolated Universe and the prevalence of a unidirectional time flow (i.e., the so-called  arrow of time).
    Under our interpretation, general relativity and quantum  theory have
    complementary time treatments given that they emerge from a
    theory of a more fundamental nature.  To model an isolated Universe, we use
    the Wheeler-DeWitt equation and then apply  the Page-Wootters method  
    of recovering time. It is argued that, if the recovery of an experience of time is  
    indeed viable in this framework,  interactions and quantum  
    entanglement  are  both essential features, even though the former is normally an afterthought or altogether dismissed. As for the one-way arrow of time,   this is, from our perspective, a consequence of including the aforementioned   interactions.
    But underlying our interpretation, and pretty much all others, is the  
    necessity for causality. It is this fundamental tenet which accounts for our experience of time but yet can only be postulated. Our conclusion is that  the ‘problem of causality’ is what  should be the focal point of future
    investigations.
\end{abstract}

\section{Introduction}

    The supposed ‘problem of time’  is an ongoing  discussion in both  
    physics and philosophy. Many different questions are raised under this encompassing topic, leading to disagreement amongst the wide range of contributors. Part of the trouble is the frequent use of buzzwords in lieu of clear-cut definitions. In order to avoid this pitfall, we will attempt to define any term that is used without relying on vague phrases. For examples of previous discussions on this topic, see \cite{time1,time2,time3,time4}.
    \par
    While some of these previous discussion present compelling arguments,
	there is no consensus as of yet. Pinning down a definition of time  alone is troublesome, as pointed out in \cite{norton}, amongst others.
	We start then by identifying what we believe to be the simplest possible definition of time, as well as any associated ingredients that are required to recover our experience of time. With these definitions in hand,  we  will assess three of the aspects of the problem of time. These are, one, the difference in how  time is treated in quantum mechanics and general relativity, two, the phenomenon of a one-way arrow of time and, three, the recovery of time in an isolated and therefore timeless Universe.

\subsection{Time and our experience of it}    
\label{def}
    
    \paragraph{Time}
     The task of defining  time presents one with  a problem  
since it cannot be accessed directly. As with space, our knowledge of  
time comes about from
studying the behavior of physical objects rather than directly measuring time
(or space) itself. Since the main interest
here is in describing the experience of time, we will avoid the debate about
the potential existence of time and space as independent
entities but, rather, focus on describing time as it affects physical systems.
(But for a summary of this debate, see, e.g., \cite{daiton}.)
We do note that the physical
existence of an extended object ---  even for one that is not moving --- would
seem to imply that space is required to provide a physical meaning to the
spatial dimensions of the object.~\footnote{As definitions of existence can lead to
many philosophical issues, we will rely here on a  notion of `physical being' that is similar
to the Parmenidean view of `what is' as  described in, for example, \cite{popper}.
More nuanced definitions can be found in, for example, \cite{phil}.}
If this is true, one might argue, by extension, that time should likewise be required, 
a point which will be elaborated on later.

\par
    To ensure that any features which are necessary
for the experience of time  do not
  become hidden assumptions, we begin with  a very basic observation:
The experience of time is one of change. In particular, we regard the `configuration' as the feature of a physical system
  that undergoes change.
By configuration,  we really mean the state –- which is defined by all the
properties that the system might have at any given moment –- but  have
tweaked the  terminology as a reminder that
the Schr{\"{o}dinger picture is then assumed for   
the quantum case.
Given a physical system, which can be constructed from one or more parts,
‘change’  indicates  the process by  which the system transitions from  
  any one configuration to another. For example,  two configurations  
of a cup could be
‘in midair’ and ‘scattered in pieces on the floor’ .
The continual change in a system’s configuration leads to the  
experience of time but, as of yet,
there is no compelling reason to insist that the configuration of any  
system {\em must} change.
  The implication is that the most basic definition of time lacks such  
a feature.
With that  said, it is important to distinguish between two related  
but still  different situations where a system does {\em not} change:

On one hand, a system may have no functional dependence on time and therefore
remains in
the same configuration regardless of the value of the time coordinate.~\footnote
{The debate over whether time has independent existence often separates `time' as a thing-in-itself from `coordinate time' as a relational concept which relates physical systems. For a summary of the difference, see, for example, \cite{daiton}. The discussion presented here, however, is not concerned with this debate and so no such distinction of coordinate time from `independent' time is made.}
Such a system is normally said to be static; while it might exist in multiple moments of time, its configuration is the same in all. A stationary system would also fit this bill given that the transitions between configurations have been suitably coarse grained.
On the other hand, a system with access to only one moment of time would also be incapable of change since it would have no secondary moment to `move' into. We will call such a system `frozen' to capture the idea that it is stuck in the one moment it happens to occupy. It could be argued that a frozen system can be functionally dependent on time as its configuration may depend on being stuck in {\em this} particular moment instead of {\em that} one. 
A `frozen' system could then be considered to be {\em a priori} capable of change, which it would experience if it had access to more than one moment of time and was compelled to `move' to a second moment. But, lacking these attributes, a frozen system, just like  a static system, is unable to evolve.~\footnote{
This view of frozen systems shares similarities with  the Block (or timeless) Universe
perspective. The latter is discussed in Section~\ref{dis} and Appendix~\ref{b}.}

But, on still another hand,  one might ask if a non-evolving  system  
  needs to
to rely on time at all in order to exist. In other words, could an 
existent system 
have  no opportunity for change as a strict matter of principle?
  If it could,  
this would provide
us with a  third distinguishable situation. Nevertheless,  
a compelling argument for the necessity of time in describing physical existence goes as follows:
Since  the configuration  is what is  affected as  
a system moves
through time,  one can reasonably assume that the configuration
is what should be influenced by the removal of time. (However, see
\cite{daiton} for a contrary opinion.) 
Then,  without the notion of a moment  to ensure that a single  
configuration has been selected out of the many possibilities, the  
system could be
viewed as  occupying  any one of its configurations. But this  scenario  
is logically no different than
claiming that
the system is occupying all possible configurations. 
We would then
argue that such a situation deprives the system of its physical existence and
conclude that time is indeed necessary, regardless of any functional dependence on a parameter of time.~\footnote{This suggestion bears some similarity to the perdurantist view, a defense of which can be found in \cite{sider}. There it
is argued that any object necessarily has a temporal aspect which must be taken into account, albeit with  a different line of reasoning.}

    \par
    It is worth pointing out that  depriving a system of the compulsion  
to change tends to put time and space on more even ground. This is  
because a basic definition of space might be that it is a feature of  
the Universe which allows movement to occur but does not include the  
compulsion to move.
Regardless of the impetus, whether it be the most basic definition of  
time or its unification with space,
we will similarly define time  as a feature of the Universe that  
allows change to occur but does not compel it. Given one's own  
understanding of the experience of time, this definition immediately  
begs the
question: what does provide the compulsion?

\paragraph{Causality}

We still need to  account for the continual change 
from one configuration to the next, and in a manner which is consistent with the
experience of physical systems. The described process can be attributed to the  principle of causality which, in most discussions, is either 
formally postulated or taken for granted to allow for an interpretation of physical theories.~\footnote
{We recognize that the status of causality is a contentious issue in the philosophy of physics and do not claim to present a proof of the principle. Rather, we are acknowledging causality's role in our current descriptions of the experience of time even as an {\em ad hoc} principle.}
Since 
the meaning of causality has many different variations, let us first clarify
the definition to be used here. What we have in mind
is similar to the viewpoint of \cite{pearle}, which talks in terms of  `interventions' acting on systems
to induce change. More to the point, causality will be taken to represent a process by which
an external influence, the cause, compels a system to transition from
one configuration to a different configuration, the effect.~\footnote
{As pointed out in \cite{pearle}, this suggests that causality cannot have meaning for a closed system for which there can be no outside influence. This concern will be addressed in Section~\ref{dis}.}
 Because one of the configurations 
is a consequence of the transitioning of the other, there is a natural
precedence for the configurations, but this is not an inherently temporal ordering until causality is
combined with a suitable definition of time, like the one above. 
Once this step is taken, 
the ordering in time can be set so that the cause  always precedes the event. 
A series of configurations  which are strung together with the requisite 
ordering provides the desired picture of a system moving through time.
 This closely resembles the account of time as
a process of ‘becoming’, which was historically first suggested by Heraclitus in
ancient Greece and, more modernly, by ({\em e.g.}) Whitehead \cite{white} and Prigogine \cite{prig}. 

\par
It should be stressed that this definition of causality does not 
implicitly include any notion of determinism. Although determinism and causality
are often conflated, and many authors  disagree that they can be separated, here the distinction is maintained. The difference can be seen by considering 
the case where a cause may have several possible effects, as is evident in stochastic theories
and, of course, in quantum mechanics. (See, {\em e.g.}, \cite{QMcaus} for further elaboration.)
\par
In spite of our claims  about a causal ordering, the direction of the time flow remains
ambiguous. To understand why, let us to return to the example of a falling cup.
There, it is natural to identify gravity as the external influence or cause and the
transition from ‘in midair’ to  ‘scattered in pieces on the floor’ as the effect.
This picture makes the ordering of the configurations clear, but it is only natural
because an observer would rarely (actually, never) see  
the reverse ordering of configurations. As is well known, the equations of motion
in most physical theories are time-reversible invariant, and Newton's equations
of motion are no exception. And it is just as well known that, with the possible exception
of the collapse the wavefunction, physics  should not and  does not require  a conscious observer
to operate. Meaning that one could just as easily say that some unknown agent caused
the plate to reassemble and project upwards. Who is to say what is the correct interpretation?
This same logic can be extended to a  chain of configurations: the sequence of the events
remains clear but the labels of
‘cause’ and ‘effect’ can be arbitrarily assigned to either end of the chain. 
It then appears that time reversibility  undermines the utility of causality in determining  relationships between configurations. (This same observation is relevant to `timeless'  models of the Universe; see  Section~\ref{dis}.) 
\par
The discussion above does, however, overlook thermodynamics, which identifies certain processes as being
irreversible: those for which the entropy has increased. In other words, the 
 second law of thermodynamics, which states that the entropy of a closed system can never decrease, enforces irreversibility and, with it, a unique direction in time.  In the previous example, it is quite clear that the smashing of the plates is just such an irreversible process. 

Before the thermodynamic argument can be accepted, there are (at least) two counter-arguments to consider: thermodynamics is not a fundamental theory, which makes it
capable of hiding the intrinsic reversibility of physics, {\em and} there is no obvious reason why this so-called 
thermodynamic arrow of time should align itself with the cosmological time,
which is taken to be the time experienced by a (typically large) group of gravitationally bound systems as a whole. 
But it will later be counter-counter-argued that irreversible processes are absolutely necessary if time is to emerge in an otherwise timeless (or isolated) Universe
and that these very same processes are what accounts for the direction of cosmological time.~\footnote{The importance of irreversible processes in the context of time evolution has been advocated by others;  for example,  Prigogine 
\cite{prig} (also see \cite{albert,ellis} for different perspectives).}   And, because the time evolution is itself emergent  and not fundamental, it is no longer subject to the same  symmetry properties of the underlying theory.

 \subsection{The `problem of time'}

    To summarize the discussion so far:  Time (by itself) allows systems to change,   causality compels them to change and, under the assumption of an isolated Universe,  irreversible processes pick out a unique direction for change to occur in. Let us now return to the aforementioned three problems of time and briefly preview our proposed resolutions.
\par    
Our first task will be to address the different treatments of time in  quantum mechanics and general relativity. As will be explained, there is reason to believe  
that these two time treatments descend from a common theory. To be clear, this common theory ---  
for which the putative fundamental theory is its antecedent ---
 does not have to contain time as its normally understood; a template will suffice. As for our reasoning, this is based on three observations: quantum mechanics is a descendant of
quantum field theory, both quantum field theory and general relativity have a spacetime metric, and the Minkowski metric of quantum field theory should be regarded as a limiting case of a generic class of metrics and not merely a background structure. The very last claim will be shown to be a consequence of quantum field theory having no global symmetries, and it means that there is fundamentally no difference between the metrics ---  and therefore time parameters --- of 
the two theories in question. 
 Here, we are not claiming to offer a conclusive description of a `quantized time' as it might appear in a theory of quantum gravity. Rather, we are presenting evidence of similarities between the two treatments of time, supporting the notion that they do in fact arise from a more fundamental source. 
\par
Now what about the arrow of time?  As already discussed, our resolution of the timeless Universe problem (see below) also provides a built-in resolution to the arrow-of-time quandary, as it ensures that the cosmological times and thermodynamic time are in alignment. 
The central point is that the emergence of time requires irreversible processes as a matter of principle, and these provide a natural direction for the time flow. There is, however, another brand of time  
to consider; namely, psychological time. This typically refers to time as it is experienced by conscious beings who have an innate ability to remember the past but never the future. Although the argument that directly connects psychological time to its thermodynamic counterpart is well known ({\em e.g.}, \cite{ar1,ar2}), we will summarize  it here;  both for the sake of completeness and because  similar reasoning is used later in the paper. Briefly, the storage of  a new memory first requires the erasure of an old one. But the latter comes at the cost of an increase in entropy due to an associated loss of heat \cite{lan}.
It is often pointed out, by way of analogy, that this process is functionally no different than adding a new bit of information to a computer. What will be important to us is that the same can be said about any physical system that is capable of storing information, even if it is inanimate. For further discussion, see Section \ref{tt}. 
  \par
    The last of the three issues is also the most involved: The problem  of providing time to the Universe if it is truly isolated.  The Universe would then have to  be in a timeless state because, simply put, it  already contains `everything'. The real point though is the conservation of energy, which  must be in effect for any closed system. For a closed system without gravity,  whether classical or quantum, one is then free to set the on-shell value of the  Hamiltonian to zero by adding a constant. If gravity is included, one no longer has this freedom, but general relativity handles it automatically through its constraint equations. And, deprived of a Hamiltonian (as far as physical solutions of the field equations are concerned), time evolution is impossible.
To be clear, a particle in a well, for example, is in a static state and not a timeless one, because time is still provided by its external environment. But there can be no such  environment to rely on when the Universe exists in isolation.~\footnote{Our discussion does not preclude parallel universes nor multiverse theories  but instead assumes that one member of such an  ensemble cannot influence any  other.}

\par
 The model of an isolated Universe is captured by the Wheeler--DeWitt equation \cite{wdw}, which 
elevates general relativity into the quantum regime (more accurately, into the realm of semiclassical physics).   
We are not claiming (nor disputing) that this equation provides an accurate depiction of reality, 
but it does exhibit the essential feature of timelessness. Page and Wootters famously proposed a method of recovering time in the Wheeler--DeWitt framework; the  basic idea is to allow a subsystem of the Universe to serve as a clock for the remainder \cite{paw}.
The delineation of the subsystems is, at first glance, arbitrary under the proviso that the systems are maximally entangled (in the quantum sense) and  weakly interacting. In subsequent descriptions of this method, the interactions are typically deactivated, at least as a limiting case. Our own investigations led to the conclusion that the  interactions cannot be arbitrarily small and still allow for an adequate description of time \cite{us1,us2}. Moreover, as one subsystem is effectively measuring the other, the interactions are necessarily irreversible. And so what we have is an emergent cosmological time with a built-in arrow of time that automatically  aligns with the thermodynamic arrow and then, vicariously, with the psychological arrow. (Similar links between general relativity and thermodynamics are touched upon in    
Section~\ref{dis}.) 
\par
What is still missing, however, is a mechanism that explains how either of the
two subsystems can transfer from one of their configurations into the next. In other words, what is still lacking  is an explanation of causality. 
Since causality is, to the best of our knowledge, always introduced as a postulate, a complete `theory of everything' may be what is needed to pinpoint its origin. In the   meanwhile, our suggestion would be to change the focus of future investigations from the problem of time to the mystery of causality, as 
it is the latter that is ultimately responsible for the experience of time in the Universe.
\par    
    The remainder of the discourse is organized as follows: Section~\ref{qm-gr} presents our argument that time in quantum mechanics and general relativity should not be regarded as independent
entities, and that their respective time treatments can be traced to a common theory.
    Section~\ref{tt} recalls  how time can emerge in an otherwise timeless Universe and then shows how the requisite inclusion of interactions forms a link between the cosmological and thermodynamic time arrows.
    Section~\ref{dis} presents additional discussion regarding the problem of time in the context of our 
findings and d conclusions.
    A supporting description of the Page--Wootters method is provided in Appendix~\ref{a}.
    Several possible objections to our conclusions are discussed in Appendix~\ref{b}, including the clock ambiguity~\cite{amb}, the Block Universe, and the inclusion of other universes. 
    
\section{Time in quantum mechanics and general relativity}
\label{qm-gr}

     The aim of this section is to reconcile the differing views of time which arise  in  different theories of physics. Historically speaking, the first physical definition of time appeared in Newtonian mechanics, and so we  begin there.

    \subsection*{Time in the classics}

    Time, as described by Newton, was an ``absolute'' quantity which existed outside and independently of physical systems \cite{newt}. As such, it could not be directly measured but only described as a relative quantity between events, where an event represents  a change in  the configuration of system and is quantified through the classical equations of motion. After establishing a coordinate system, one 
can use these equations to assign each event with a location in both space
$\vec{r}$ and a time $t$.
    If different coordinate systems are moving  relative to one another at a constant speeds, 
each one can be 
viewed as  the reference frame for a co-moving observer.~\footnote{Our use of terms such as  `observer' and  `perspective' should not be taken to imply a conscious experimenter.   We will be explicit whenever such an observer is required to make a point.}
Moreover, any two of these inertial frames can be related though a Galilean 
transformation
of their  spatial coordinates. Note though that  the time axes of two such reference frames can never be moving relative  to one another, as $t$ applies globally to all frames (up to an arbitrary choice for  $t=0$),
 making the  difference between time and space  quite  apparent. And so  time in Newtonian mechanics, 
even as a parameter describing relations between events, remains external to the events. We will refer this  parametrization  as   `external time'. 
    
  The  external time $t$ requires a further assumption if it is to describe dynamics.  Since the external time  exists outside 
of the physical arena, there is no inherent motivation for a series of events to be ordered by, say,  assigning each with  a  particular value of $t$. 
   This  mathematical framework, at best,  only provides  an elementary  notion of time as already described in the introductory section.
    The principle of causality is still required to motivate a recognizable order for any series of  events. Suppose that each configuration of  a system  is represented as a frame from a film reel. Without causal relationships, the frames may be shuffled into any arbitrary order without contradicting the mathematical rules of Newtonian mechanics but also without producing a recognizable  description  of reality. 
  Only when causality is enforced (or postulated) will  the frames be restricted to a specific order. This sequence accounts for the `moving' from one configuration of the system to another; what is better known as dynamics. 
    
   But, as was unknown to Newton, his mechanics emerges from special relativity in the limit $\;c\rightarrow\infty\;$ or, alternatively, from quantum mechanics in the limit 
$\;\hbar\rightarrow 0\;$, both of which are illustrated later in Figure~\ref{fig1}. Of course, these two theories are themselves limiting cases of  theories that are even more diverse in their scope.  Meaning that external time  and all its trappings should be viewed with a healthy dose of skepticism.  With this in mind, the discussion will advance further up  the `ladder
of fundamentally', starting with an examination  of time in special relativity.

    \subsection*{Time in the relativistic theories}
    
   If the  Newtonian treatment of time is an artifact of its treatment in 
special relativity, how does the more fundamental notion of time differ?
   
    Many discussions on special relativity ({\em e.g.}, Chapter 2 of \cite{gr}) begin with the  postulates that  the laws of physics and  the speed of light are the same in
all  inertial reference frames. Using these requirements, 
  one finds that  a pair of such frames  can now be related by what are known as Lorentz transformations.~\footnote{Lorentz transformations are described by 
rotations in space  and/or  boosts (rotations in four-dimensional Euclidean space). The complete set of transformations are
the  Poincar{\'e} transformations, which also include translations.} These transformations are fundamentally different  from the Galilean 
transformations of Newtonian physics in that  time and space can both undergo  transformations and can, indeed, even  become mixed.
To assure that the laws of physics remain intact, physical quantities are required to transform covariantly under Lorentz transformations.
On this basis, one can rather construct the theory of special relativity
by enforcing Lorentz covariance where appropriate.
Importantly, there is a smaller class of quantities that, just like the speed of light, must remain unchanged under Lorentz transformations;
these  being the Lorentz invariants (typically scalar quantities, but
see below for a notable exception). 
As the set of all Lorentz transformations forms a group,
a Lorentz-invariant quantity is a textbook example of a global symmetry.

 An important  example of a Lorentz-invariant scalar quantity is proper time; 
this being the amount of elapsed time between two points in spacetime  
as would be measured by a clock moving along  
a strictly timelike path (see below for a definition). 
A path-independent way of measuring proper time 
 can be obtained by introducing the Minkowski  metric tensor. Although not a scalar,
the Minkowski metric is  yet another object that is invariant under Lorentz transformations. It is also the uniquely coordinate-independent form of
a more general tensor.
 In general,  given a spacetime manifold (flat or otherwise), the associated metric tensor provides
a means of measuring the  square of the  `distance'  between any two points in the manifold;  this being the same, up to sign,  as measuring the square of
the proper time interval.  

   In discussions about spacetime geometry, the notions of space and time can become blurred because of their aforementioned mixing transformations. The proper time is one way of maintaining a distinction. Another,  more geometric way
is provided by  null cones; this being an  object  whose surface is defined by the path of the light rays emanating from its apex. As light is the fastest-moving object from any observer's perspective, the null cone serves as a means for delineating between timelike paths --- those staying within the null cone --- and spacelike --- those passing through its  outer surface. The boundary of the cone itself contains
strictly null (lightlike) paths.  For a more graphic description, see Figure~\ref{sr}.
 It should also be kept in mind that the notion and utility of both  proper time
and light cones persists for more general spacetime geometries. 
        \begin{figure}[ht]
    \centering
      \includegraphics[width=0.8\textwidth]{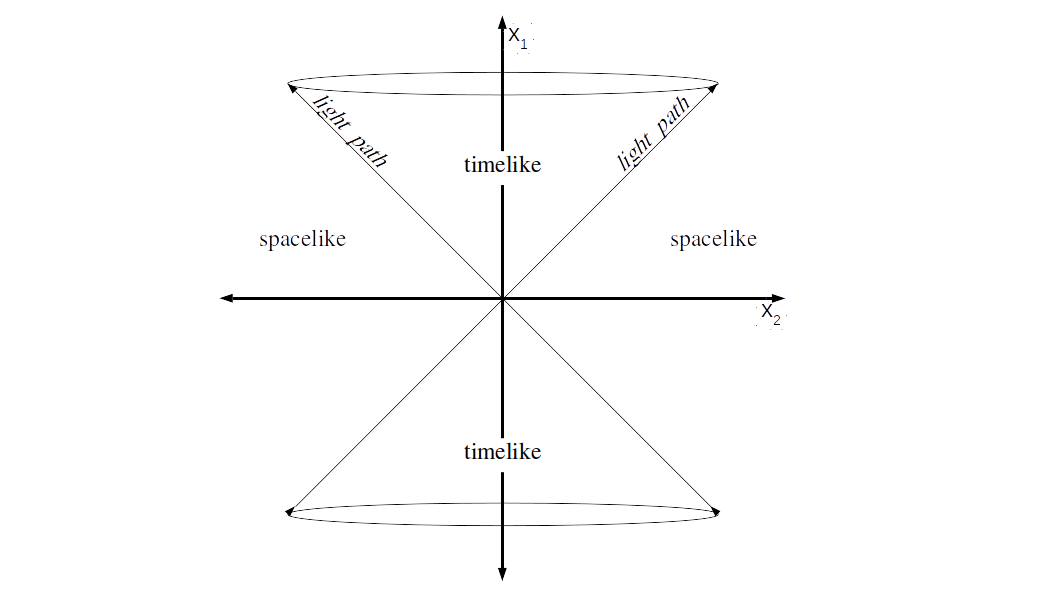}
    \caption{\em A schematic diagram of a  null cone,  whose outer surface 
describes the path of light rays in spacetime.}
      \label{sr}
  \end{figure}
  
 It is clear that, at least in flat spacetime,  the null cones identify just one of the four spacetime dimensions  as 
temporal. But time, as it is experienced in Nature, is not included in this setup as long as the notion of  causality is absent. Same as for
Newtonian physics,  special relativity does not necessarily produce recognizable dynamics. As per the previously described film analogy, special relativity might allow for the  frames to be fixed together but it fails, on its own, to
ensure that the sequence of frames will be arranged in any particular order.  
   
    Special relativity is, of course, the weak-gravitational or $\;G_N\to 0\;$ limit of its more fundamental description, the theory of  general  relativity. But translating between theories is much more involved than applying a simple limit.
For one thing, in the general theory, inertial reference frames are no longer favored over their non-inertial (or accelerating) counterparts. For another,  as the effects of gravity are `turned on', the spacetime geometry becomes curved
and, as a result, the metric tensor will  no longer be expressible in a constant form nor will it be invariant for an arbitrary Lorentz transformation. In fact, because two reference frames can
have a relative acceleration between them, Lorentz transformations are no longer sufficient. 
 Rather, diffeomorphisms (or generic coordinate transformations) are  now required to map from  one reference frame to another. Meaning that physical quantities are now
 required to transform covariantly under diffeomorphims and there is  a privileged class of quantities that
are  diffeomorphism invariant. The spacetime  metric itself is, itself,  not such an invariant 
and will generally  change according to its position in spacetime but always in just the right way to ensure that scalar quantities, like proper time, remain
 diffeomorphism invariant.

And so, rather than saying that special relativity is a weakly gravitating limit of general relativity, one might  be more accurate in claiming that general relativity
comes about by breaking a global symmetry --- that of Lorentz-invariant scalars --- into 
a  local ({\em i.e.}, coordinate-dependent or gauged)
symmetry --- that of diffeomorphism-invariant scalars. 
From this point of view, 
    the Minkowski metric is not so much the  limiting case for a flat spacetime 
geometry  but more like the metric whose associated gauge field can be fixed so as to 
trivially vanish. As the gauge field holds all the coordinate dependence,
the Minkowski metric maintains  a `hidden' dependence on the spacetime coordinates.  This dependence has, however, as a matter of choice, simply been gauged away.

As already mentioned,    null cones remain a part of general relativity but are generally deformed away from their flat-space geometries. An illustration of this is
provided   
n Figure~\ref{gr}. Other notions like timelike paths and proper time 
are similarly intact but, albeit, more challenging to (respectively) identify and calculate.
    \begin{figure}[ht]
    \centering
      \includegraphics[width=0.8\textwidth]{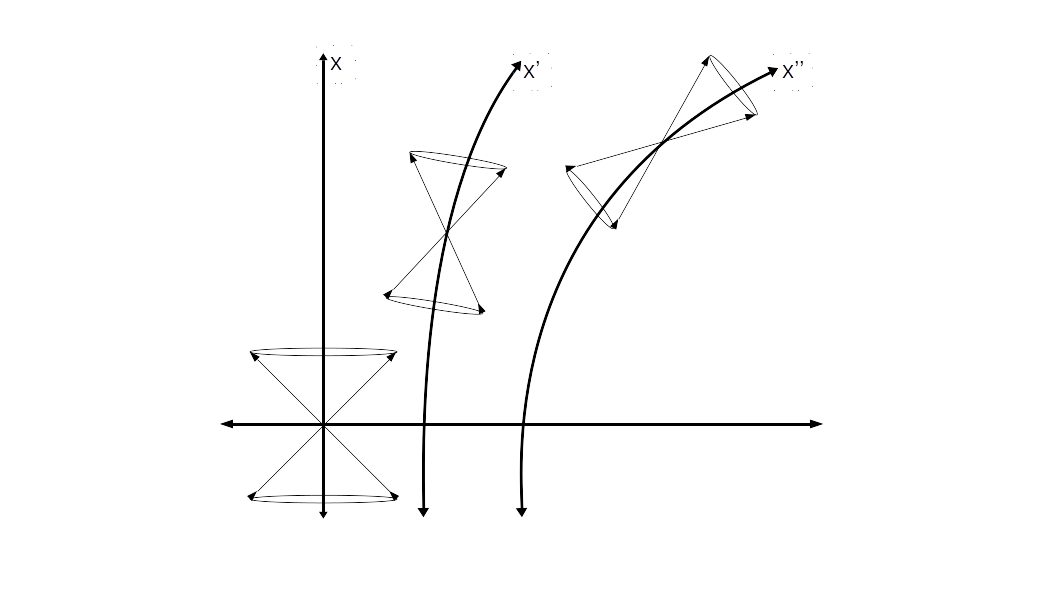}
    \caption{\em A schematic diagram of null cones `warped' by the curvature of spacetime.}
      \label{gr}
  \end{figure}  
    However, ordering the configurations  of a system into a physically 
significant sequence once again requires the postulation of  causality.
 This is contrary to the idea that time, as it is experienced in Nature, emerges from the framework of general relativity as is sometimes claimed. 
    
    It can be  anticipated that general relativity itself emerges from the fundamental theory or,  perhaps, from an intermediary thereof  ({\em e.g.}, string theory \cite{str}). 
Before further speculating on this possibility and its relevance to the emergence of time, let us first
 turn to another, seemingly unrelated  series of physical theories. 
    \subsection*{Time in the quantum realm}
    
   Although the transition is not well defined, classical Newtonian behavior can be expected to emerge from quantum mechanics  in the limit  $\;\hbar\rightarrow 0\;$. 
But the transition of time does appear to go smoothly, as  quantum mechanics, 
like its  Newtonian counterpart, utilizes a global time parameter  for the purpose of
ordering  the configurations of  any given system.
    In the quantum case, the necessity for such an external time line can be traced to the theory's  prohibition on quantum time  operators, which applies even as a
strict  matter of  principle. 
The argument against a time operator is well known but is still worth reviewing:
Any  time operator would necessarily be conjugate to a  energy operator ({\em i.e.}, a  Hamiltonian), as follows from the relation  $\;[E][\Delta t]=[\hbar]\;$. Now let us suppose that a time  operator $\hat{t}$ does indeed exist. 
Then one could construct a unitary operator $\;\hat{U}=e^{\pm\;i\;\hat{t}\;dE}\;$ which 
acts to translate states along the energy spectrum,
 $\;\hat{U}\ket{E}\rightarrow\ket{E+dE}\;$. Such a translation could be applied indefinitely, projecting the system into a state of arbitrarily negative energy and, thus,  removing any notion of a stable vacuum state. 
    
  Things finally get  interesting when the  more fundamental theory of
quantum fields is brought to the fore.
  As the  synthesis of quantum mechanics and special relativity, quantum
field theory can be expected to limit to quantum mechanics as 
$\;c\rightarrow\infty\;$. Note, though, that this limit, just like the quantum-to-classical case, 
lacks  a well-defined transition as it would entail one to (somehow)
`deactivate' the so-called second quantization of the field-less theory.
But the takeaway point should be 
that the description of time in quantum field theory
must mimic the time treatment  of special relativity if the field theory is to 
maintain its integrity. Meaning that quantum field theory, just like special relativity
and all the others, depends on the postulation of causality if it is to describe
the experience of time.
    
    \subsection*{Relating time across theories}

    As  illustrated in Figure~\ref{fig1} and  discussed above,  there  are  
two ways to reach classical Newtonian physics from the  more fundamental theories. Importantly, these are one-way paths; Newtonian mechanics can inherit time from 
either pathway, but one cannot go up along one path and then backtrack down the other in order to connect general relativity 
to quantum field theory.   But this pair could still meet at the opposite end in an even more fundamental theory, as depicted in the figure. Although this is our expectation, we will proceed to argue that the
two paths depend on the same basic notion of time, irrespective of whether or not they meet at  the `top'.

\begin{figure}[ht]
    \centering
      \includegraphics[width=0.8\textwidth]{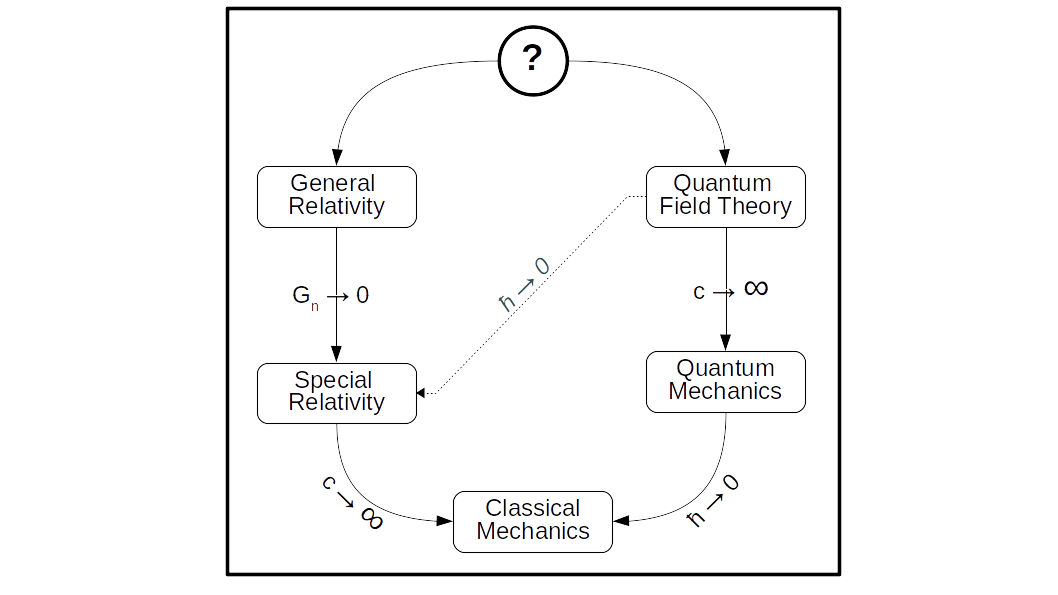}
    \caption{\em Schematic diagram showing direct relations between theories in physics where `?' stands in place of all the as yet undetermined theories: a fundamental theory and any intermediary versions thereof.}
     \label{fig1}
  \end{figure}

Quantum field theory and general relativity are commonly  viewed
 as disconnected entities in `theory space'. But yet they both 
include special relativity as a limiting case, which opens up the possibility of a hidden connection between the two theories.  However, given the usual state of affairs, this would be  a naive connection at best as  the two theories provide
much different  interpretations  of
the spacetime metric which special relativity ultimately inherits.  In the case of general relativity, the metric
tensor is a physical field that  describes the geometric and causal structure
of spacetime. And so, from this point of view, the Minkowski metric of special relativity can be interpreted as  the flat-spacetime limit 
of a physical object.

Meanwhile, the Minkowski
metric of quantum field theory is generally regarded as a mathematical construct
and not a real physical object. This description falls  in line with that  of the gauge fields in classical electrodynamics, 
which seems sensible for a classical theory but
somewhat contradictory for a theory of quantum (gauge) fields. We will  argue
 next that this interpretation of the Minkowski metric is indeed misguided and that, just like in the flat-spacetime  limit of general relativity, this tensor is  harboring  a hidden dependence on the coordinates of spacetime, making it 
a physically real field. Note, though, that any such dependence must  remain hidden and undetectable so as not to jeopardize the requisite Lorentz invariance of quantum field
theory nor the closely related theorem of Weinberg and Witten \cite{ww}.

Our first argument is undoubtedly the simplest one: Quantum field theories 
can be  expected,
on very  general grounds, to not have any unbroken global symmetries   \cite{wit}, and we 
see no good reason for  Lorentz invariance to  be exempt from
this `policy' (given that the effects of it breaking  remain  hidden as just discussed). The relevant point here is the inevitable breaking of classical 
 symmetries after the theory has been suitably quantized and
renormalized; the so-called quantum anomalies  \cite{wit1}.

  A second  argument comes   from string theory, which is also void of global symmetries \cite{wit2}.  Even if string theory does not accurately depict reality, it is the only known
 self-consistent theory which accounts for both general relativity and quantum field theory and so it can be used as an indicator of what a more fundamental theory might look like.~\footnote{String theory is certainly more fundamental but we are not asserting that string theory is {\em the} fundamental theory. In fact,
we would argue against such a suggestion
(see, {\em e.g.}, \cite{st} for a relevant discussion.)}

Our  third argument goes as follows: Suppose that there was a more fundamental theory  which contains both general relativity and quantum
field theory.  We would then expect all of its  observable and emergent features --- including proper time, 
null cones and, by extension,  the metric ---  to be subject to 
the effects of quantum fluctuations. 
Such fluctuations can not be  prohibited from depending on the coordinates
of spacetime.

Let us, finally, recall a standard argument that is routinely used against global symmetries: One  considers particles that are forever lost inside of
 a black hole, which implies the  breaking of  globally conserved quantities such as the  baryon number \cite{beken}. Note that the process of  black hole evaporation process cannot resolve this situation because  the emitted radiation is dominated by 
massless  particles \cite{page} and any such  particle is incapable of
carrying  a baryon number. 

The two middle arguments imply that there is some intermediary  theory or,
possibly, the fundamental theory itself from which both general relativity and quantum field theory are emergent ({\em cf}, Figure~\ref{fig1}). Although not strictly necessary (as the other two arguments would suffice),  this
would be the natural expectation  if one accepts our assertion  that the two
theories do indeed host similar time treatments. But, to be clear, we are not 
suggesting  that a recognizable  notion of time has to exist in all theories up to 
and including the fundamental
theory, but only that  some `blueprint' for time is provided at a more fundamental level.
This feature will be referred to as `elemental time' later on in the paper.
 
To summarize, our contention is  that all currently accepted  theories
are really talking about the same basic notion  of time but, by the same token,
all require that causality be postulated. 
 
\section{The thermodynamic arrow and the timeless Universe}
\label{tt}

    Whereas  the thermodynamic arrow of time points in the direction of increasing entropy, the closely aligned psychological arrow 
is taken to represent time as perceived by  a conscious observer recording experiences in her 
memory. These arrows are  well known but what may not be is that the concept of a psychological arrow can be extended beyond 
 conscious observers such as ourselves and  even beyond computers.
The same idea applies just as well to  any physical system 
that has changed its configuration  as the  result of an irreversible 
interaction. 
The point is that any such change can be viewed as  the act of `recording information' in the sense that one configuration (or `memory') of the system  is erased and replaced by a new one. The constraint of irreversibility ensures that the previous configuration can only be restored with the sacrifice of a new one
 and accounts for the 
alignment with the thermodynamic  arrow.
The latter because such interactions produce heat and then,  by virtue of the Clausius inequality, increase entropy. What we intend to show here is that this arrow emerges as a  consequence of the Page--Wootters methodology \cite{paw}, but only if one also insists on a  physically meaningful notion of time.
 
    \par
   The phenomenon of time depends on one's interpretation of  the Universe.
If the Universe is regarded as a totally isolated system, as presumed here, then the phenomenon of time cannot be imported from the
 `outside' --- it must rather emerge from within. This isolated model of
the Universe is captured by the Wheeler--DeWitt equation \cite{wdw},
    \begin{equation}
    \label{1}
     \hat{H}\ket{\Psi}\;=\;0\;,
    \end{equation}
    where $\hat{H}$ is the Hamiltonian constraint from general relativity but  elevated to the role of a quantum operator and $\ket{\Psi}$ is the putative wavefunction of the Universe. As a closed and stable gravitating system, the total energy of the Universe is zero and unchanging. As such, the Hamiltonian annihilates all physical states; meaning that the relevant states cannot  evolve in time,  $\;e^{i\hat{H}t}\ket{\Psi}=\mathbbm{1}\ket{\Psi}\;$. The conserved energy and 
the static nature of the states  should be regarded  as a manifestation of the Universe's isolation and not as a unique feature of the  Wheeler--DeWitt description. 

    \par
   The emergence of time is problematic given that the Universe is prohibited from `fetching' any of its features from its (hypothetical) exterior. 
It is useful to compare this situation to that of isolated quantum systems in a 
more conventional setting. In standard quantum mechanics, the difficulty is overcome simply  because the isolated system exists within an environment that does experience time, and so a  notion of  time can still  be  imported from this exterior region. Not unlike a Russian doll, any isolated system can be fitted inside a larger one so as to import a notion of time,  with the  procedure repeated indefinitely until some  largest possible system is reached.
In the case of the isolated Universe, however, one is beginning the iterative procedure already at this upper maximum.

    \par
Let us now move on to the Page--Wootters solution to this conundrum.
 Those authors proposed that the Universe be divided  into two subsystems \cite{paw}: the clock $C$ and the rest of the Universe $R$. Here, we will only briefly summarize the method but have also  included a more quantitative description in Appendix~\ref{a}. One begins by  subdividing the Universe in such a way 
that $C$ and $R$ are maximally entangled and approximately isolated.  
The conjugate operator $\hat{P}_C$ to the 
clock Hamiltonian $\;\hat{H}_C={\rm Tr}_R\hat{H}\;$  can than serve as an  effective time operator for  $R$.  
The reason that this works is because the approximate isolation ensures 
that, as far as physical states are concerned,  the respective Hamiltonians are related by  $\;\hat{H}_R\approx -\hat{H}_C\;$,
 and so  the effective time parameter for $C$ (namely,
the eigenvalue of $\hat{P}_C$)    and that for $R$ are the same up to a sign convention.  The condition of maximal entanglement 
is itself necessary to ensure that the states of $R$ are indeed correlated 
with the eigenvalues of $\hat {P}_C$ in a one-to-one way (assuming no degeneracies). Any interaction effects, which would be governed by a Hamiltonian of
the form  $\;\hat{H}_I= \hat{H}-\hat{H}_C-\hat{H}_R\approx 0\;$, are 
negligible in this setup and the so-called  ideal-clock limit
implies that  $\hat{H}_I$ vanishes identically. 
    Although this treatment suffered criticism from Kuchar \cite{kuch}, the issue has since been  resolved by independent investigations \cite{cc,ll}.

    \par
  The Page--Wootters process thus restores  the notion of time in  the (otherwise) timeless Universe but not necessarily in a useful way.
 Given that the clock and its complement are roughly equal in size,
$\;\dim{C}\sim\dim{R}\;$,~\footnote{The consequences of weakening this assumption are discussed in Appendix~\ref{b}.} 
the emergent time parameter
can  only provide a time ordering for $R$ as a whole; it cannot do so for an arbitrary subsystem of  $R$. 
    \begin{figure}[ht]
    \centering
      \includegraphics[width=0.8\textwidth]{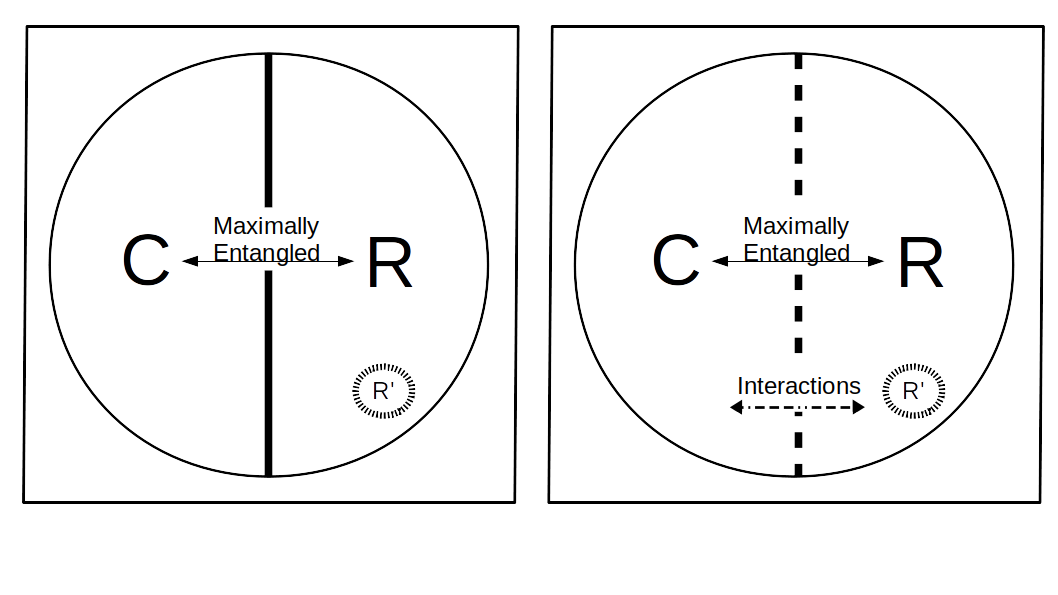}
    \caption{\em A representation of the Hilbert space of the Universe as used in the Page-Wootters method with  clock system $C$ maximally entangled with the rest of the Universe $R$. The left image shows an isolated clock system. The right image shows a clock which is allowed to interact with an arbitrary subsystem $R'$ contained in $R$.}
      \label{fig2}
    \end{figure} 

  This point underlying this last claim is illustrated in Figure~\ref{fig2}, where one can see that
    $R$  has to `disperse' the mutual entanglement throughout a much larger 
region of Hilbert space than that occupied by a smaller subsystem,
$\;R'\ll R\;$. That is, $R'$ can have only  partial `knowledge' of the mutual entanglement  since this smaller subsystem is not,
by itself, strongly entangled with $C$. So that, as far as $R'$ is concerned,
an essential ingredient of the
Page--Wootter's framework is  absent.   This appears to limit the  utility of their 
method to the case of a  two-system picture. However, a way of circumventing  this obstacle  is still possible.
    
In some previous investigations \cite{us1,us2}, we have made the case that $C$ cannot be completely isolated from $R$ if it is to efficiently  provide a  description of time.
 This conclusion depended on  some specific physical examples, but the 
analytic findings are supported by a more general argument:  
As long as $R'$ is 
continuously interacting  with $C$, its state (or configuration) will remain
correlated with the states of $C$,  which are in turn still correlated
with the eigenvalues of $\hat{P}_{C}$. Hence, these very same eigenvalues can  
also serve as
the time parameter for $R'$. 

It is worth  augmenting  this argument with a particular example ({\em cf}, 
Figure~\ref{fig2}).
Let us suppose that the role of the clock is played by the cosmological expansion of 
the Universe as often put forth in  earlier literature; see, for example.
 \cite{vil}.~\footnote{We are assuming that there is some mechanism  and thus subsystem of the Universal Hilbert space that
is responsible for the expansion.}
    Let us also  identify $R'$ and $R''$ as a pair of `tiny' subsystems of $R$; namely, the Milky Way and some other galaxy that is close enough for there
to be a mutual attractive force. There are now
two possibilities:
 Either the gravitational pull between $R'$ and $R''$ is not strong enough to overcome the expansion of the Universe and the galaxies move apart, or  the expansion effect loses out and the two galaxies move toward one another. In both scenarios, the resultant interaction between $R'$ and $R''$ is influenced by the expansion, and so by $C$. In other words, each galaxy is continually  provided with a record
of the expansion or, effectively, with a record of the time . In this setup, the light emitted from $R''$ and captured by $R'$ (and {\em vice versa}) is what represents the `clock readings'. 
   
The expansion of the Universe has recently been adopted as a clock in a paper  by Stupar and Vedral \cite{ved2}. In that study, and in contrast to ours, the interaction effects  were regarded as  negligible to ensure that the clock is  
ideal ({\em i.e.},  a completely isolated system). Although the clock in these 
examples is only one of many choices, we would contend that it is the natural one because, in this case, the clock parameter is literally the cosmological 
time.
Moreover,  this brand of  interactions must apply to all systems
as a matter of principle  because of
the uniquely universal nature  of gravity; everything gravitates!
  
It is important to emphasize that any interaction between the various subsystems is  associated with  the `recording of information' and, thus, with an accompanying production of heat. Hence, each successive configuration of the clock, as measured by $R'$, corresponds to an increase
in entropy and so a thermodynamic arrow of time. The mutual entanglement
between $C$ and $R$ is reduced  by these same interactions, and so an arrow  pointing in the direction of decreasing entanglement can be identified as the cosmological arrow of time. In this way, we anticipate the lining up of all the relevant arrows of time: cosmological,
thermodynamic and psychological. But the elephant in the room continues to be
the need for  causality to be postulated. 

While it might be argued that the entanglement can be discarded if interactions are allowed to measure the clock directly, we briefly point out that removing the entanglement would result in a breakdown of the Page--Wootters description of time. Let us suppose that $\ket{\Psi}$ is a separable, rather than entangled state of $C$ and $R$,  which would then lead to 
\begin{equation}
 \begin{split}
      \ket{\Psi_p}\;&=\;\alpha_p(e^{-ip\hat{H}_C}\ket{\phi_p}_C)\otimes\ket{\phi_p}_R\\
      &\neq\;\alpha_p\ket{\phi_p}_C\otimes(e^{ip\hat{H}_R}\ket{\phi_p}_R)\;,
    \end{split}
\end{equation}
where $p$ is the eigenvalue of $\hat{P}_C$, $\alpha_p$ are  coefficients, and $\ket{\phi_p}_C,\ket{\phi_p}_R$ are the states of $C$ and $R$ respectively. 
Importantly, there is no intrinsic time that can be utilized  by $C$ in the absence of entanglement. If we consider $R$ as the clock from $C$'s perspective, the removal of entanglement translates into a given state of $C$ being able to `pick' any state of $R$ as its `partner', resulting in a nonsensical description of time. The interactions would only be able to recover a meaningful sense of time without entanglement if their strength could be turned up such that they significantly influence $C$. This would negate the weak interaction condition and thus cause a breakdown of the formalism's ability to describe time. There is, however, reason to think that just  such a scenario may be realized eventually. 
    \par
We have previously argued that,
 although the effects of the interactions between subsystems may be negligible to begin with, they would  apply continually and eventually reach a point when the ideal-clock limit no longer makes sense, even as a limiting case \cite{us1,us2}. The current situation, however, does not face such a crisis. As the expansion continues,  $R'$ and $R''$ will either separate to a large enough distance 
to render the interactions as truly negligible or they will merge into a single system for  which the expansion has no influence.  Of course,  either scenario prevents any further access to the states of the clock.  Extrapolating this idea to longer and longer scales,  one comes to the realization that, for the inevitable Universal state of
maximum entropy, the requisite  interactions would become utterly
 irrelevant, as no new information about the state of $C$  could ever be recorded.  In other words, the loss of the clock parameter would occur just when 
the evolution of the Universe has finally stagnated.

\section{Concluding discussions}
\label{dis}

\paragraph{The status of time:}
   Our previously stated definition of time, as the potential for change, is distinct from the {\em experience} of time. This is because real physical systems obey a compulsion to change through a particular sequence of configurations and 
not just the potential to do so. Moreover, the specific order of the sequence plays a role in that it  maintains the consistency of physical laws, which
are what  determines how to relate one configuration to the next. Static systems are defined as those existing in a single configuration and so are not functionally dependent on time. This class of systems can be distinguished  from 
those which are
functionally dependent on time but have no compulsion to change; what we have
 referred to as being frozen in time.   
    While it is not outside the norm to see the latter category labeled as `timeless' (see below),  we have argued, much to the contrary, that the physical existence of systems --- even
if fixed in this way ---  still require a moment of time just like a motionless object requires 
a position in space.  If this assertion  survives under closer scrutiny, theories which claim to remove time completely may be at a disadvantage.
 
    \paragraph{The  timeless Universe:}
The timeless description   has its roots in the eternal Universe that was  proposed by Parmenides in ancient Greece \cite{popper} 
but has  since  become associated with more modern representations.~\footnote{See, for example,  the discussion on Block-Universe models 
in Appendix~B.4.} 
    \par
    The timeless interpretation of the Wheeler--DeWitt model 
is similar to some  previous versions, except that it  also incorporates the stochastic nature of quantum mechanics. A conceptual framework for this picture has been  laid out by Barbour \cite{time2}.~\footnote{There are other applicable frameworks such as that in \cite{time4}. Here, we are trying to capture the general features which are shared by most of the timeless interpretations.} 
That author presented a view in which there is an infinite ensemble of  distinct frames, each of  which  includes  a
collection of  physical systems in a very specific set of configurations. One
could view the
ensemble of frames as  randomly `scattered' in an infinitely
large  heap. The frames might then be `lined up' to   form an ordered sequence of configurations for each physical system. If time is indeed required for physical existence, then each such frame could still represent a  moment of time.
If this is indeed the viewpoint, then  it is fully compliant with our notion of a frozen (but
not static) version of the Universe.
    
\par
    Whereas the notion of a moment might survive in this manner, the concept of change cannot. 
    An argument prohibiting change as a real physical process  was presented long ago  by McTaggart \cite{mcT}. Although disagreeing that the argument applies
in general, we find that  it works nicely in the context of 
such timeless descriptions and adapt it accordingly. 
Let us assume  a  time-frozen  Universe (as described above) and consider  
 a physical system with an infinitely large  number of different configurations,
each  of which is contained in some  `slice' of time. The argument points out that all configurations of the system can claim equal existence in any given 
slice as there is,  {\em a priori}, no rule against this. 
The way out of this nonsensical scenario would be to insist  
that each configuration is  constrained  to exist in  a single frame. 
As a result, the identity of the physical system would have to  be  separated into `parts', with each of these being an individual physical entity 
that is existing, perpetually,  in a particular frame with its single configuration.
The physical system, as the sum of these parts, could not experience change 
in any objective way,  as no single part of it ever transitions into any other. The conclusion is that any perceived change in
a timeless Universe  would have to be illusionary. This  would be  similar to the way in which  individual frames in a sequence can be used to form the illusion of a moving picture, just like in twentieth-century animation.    
   % However, given that any number of sequences of frames are possible, one still a needs to  recover the specific order which matches up with the experience of systems in Nature. 
    
\par
    To maintain this illusion of timeless change, the `memories' --- or the records of past interactions ---   that are  present in each frame may be used to 
reproduce the specific order of configurations which are observed in Nature.~\footnote{Here, we mean memories (and records as well) in the broadest terms possible. For instance, a sea-side rock might remember being eroded by the tides.}  
%To avoid anthropic arguments, we again point out that this need not apply to a conscious observer's memories, simply a record of past interactions. Any particular frame may contain a record and so provide a history. 
\footnote
{Other timeless interpretations focus on the relations between frames to describe an order, but these would  presumably also rely on a record in each frame to account for any knowledge of other frames.}
But, for any given frame in the sequence, there could  only ever be circumstantial evidence 
for the existence of (causally) older frames. Indeed,  each individual frame isolates the physical entities  within it --- along with their respective records of past interactions ---
from any and all other frames, as each member of the ensemble exists independently of the others. Meaning that such a record cannot, in fact,  
be attributed to  interactions; it can  only be related
 to the frame that hosts it.
   % The memories are then not evidence of change but, rather, only a part of a single physical system which exists independently of all others. 

And, if this state of affairs was not already problematic enough,
an explanation is still needed to account for the specific order 
of configurations which appear in Nature. 
An advocate for timelessness would  
require something like, perhaps, a many-paths approach
or an action principle  to explain 
why the illusion of change (as perceived through the ordering of the frames) is consistent with the laws of physics. What cannot, however, be relied on is
the  notion of causality, as this conceptually breaks down when confronted 
with the timeless interpretation.

As pointed out in Section \ref{def}, causality runs amiss when the labels of `cause' and `effect' can be arbitrarily assigned to either end of a process. This  would be the case in  any theory that maintains time-reversal invariance under all circumstances, which is certainly the case here, and forms part of  
Hume's concern that, in any such process,  cause and effect would each be  equally responsible for the other \cite{hume}.  But, more importantly,  without the notion  of objective  change, the motivation for connecting one configuration of a system to  any other becomes {\em ad hoc}, never mind connecting them is a causally ordered  way. What is needed for this to work would be
the inclusion of an intervening process such as an irreversible interaction. In other words, maybe change
could be an illusion but causality could never be.  

    \paragraph{Change as a real process:}
There is a school of thought that real physical change can be viewed as a process of  `becoming', aligning with ideas from  \cite{smolin,carrol}.
Our results seem to be in support of this viewpoint. For us,  
irreversible interactions serve  as  agents of change, inducing 
each configuration of a system  to transition to --- or  become --- the next one in the sequence. This picture enabled  us to recover time as an emergent phenomena,
in spite of the isolated state of the Universe. 

    \par
  The aforementioned argument of McTaggart has been used to prohibit a  
description of the Universe as a place of `becoming'  due to the inclusion of a
  present-time or  `now' moment.~\footnote{The principles of relativity
are sometimes used for the same purpose. But there are arguments that counter
such a claim. See, for example, \cite{ellis,bouton,erasmus}.}
However, the argument can only  apply to 
the Parmenides Universe and other  timeless versions for which every moment of time is to  be  regarded as existing equally ({\em i.e.}, existing simultaneously
and in perpetuity) \cite{daiton}.

This distinction between `timefullness' and timelessness reinforces the conclusion above: Treating change as a real process requires one to  dismiss the  notion that all moments of time could exist simultaneously. 
This treatment  does, however, necessarily  invoke causality
if it is to be consistent with the experience of time in the Universe.
While the description of interactions within a closed system is often said to be incompatible with causality ---  as no external influence is available to
initiate a causal chain of events ---  
there are still  arguments to the contrary; see, for instance, \cite{frisch}.

On the other hand, adopting the stance that change is an illusion, one would 
still require a mechanism that  enforces the correct  order on a sequence of configurations. There is then no added cost for postulating causality.  

\paragraph{Final thoughts:}
The recurring theme of this work is that
the basic definition of time does not appear to be an issue but the experience
of  time remains unexplained. 
  Put differently, the necessity for  causality and that it must be put in by hand is what appears to be the real `problem of time'.
 
    The notion that time in quantum mechanics and time in general relativity are unrelated might be misleading given the expectation of an underlying fundamental theory. As explained earlier, the quantum and relativistic time treatments  could plausibly be interpreted as emerging from the same source. Under this interpretation, it would not matter that the two  treatments appear different,   as 
their respective theories have different domains of applicability --- except at the very scales
where a more fundamental theory needs to be considered.

    \par
We have also proposed a means by which the experience of time can emerge
by way of  the Page--Wootters framework for clocks in an isolated Universe. 
The key new element in our proposal is the role played by irreversible interactions; not that
their inclusion  is in any way novel  but, rather, that  their  importance in the framework has been under appreciated.
It is this class of interactions that leads to the recovery of a  
Universal arrow of time; meaning that any system's sequence of configurations
will naturally line up  with the thermodynamic and psychological  arrows, and can even align  with the cosmological arrow if the expansion of the Universe adopts the role of the clock. (To be clear, it is the entanglement between the clock and the remainder which  ensures that all systems see the same set of arrows.)     
Our proposal is also indicative of  a natural connection between gravity, as per the cosmological expansion, and entropy, through the associated recording of information. Although highly speculative,  we are
tempted to suggest that this is yet another manifestation of the close-knit relation
between gravity  and entropy that has become ubiquitous in the high-energy literature. For instance, Jacobson's derivation of Einstein's equations via the first law of thermodynamics \cite{jac}, the  link between entanglement entropy and
null surfaces \cite{Ryu} that follows from the gauge--gravity duality \cite{dualities} and the recent recovery of the second law in the same holographic framework
\cite{ads}.

    If we can establish the above interpretation of time, then what still remains? Once again, all such queries  lead to the conspicuously absent explanation for causality. Without this fundamental principle, one  cannot describe any process of change in Nature. Then,  insofar as  all sensible  treatments of time 
are reliant  
on the concept of causal order, we would   like to suggest that
 the problem of time be restated as the `problem of causality'.
    By removing the clutter from the discussion, our hope is 
 that the  understanding of  time  can  advance just like time inevitably does itself.

 \section*{Acknowledgments}
          The research of AJMM received support from an NRF Incentive Funding Grant 85353 and  NRF Competitive Programme Grant 93595.  KLHB is supported by an NRF bursary through  Competitive Programme Grant 93595 and a Henderson Scholarship from Rhodes University. This work is based on the research also supported in part by the National Research
              Foundation of South Africa (Grant Numbers: 111616).
    
    \appendix
    \section{Page and Wootters}
    \label{a}
        Section \ref{tt} introduced a timeless model of the Universe, as expressed  by a pure state $\ket{\Psi}$,
along with  a Hamiltonian $\hat{H}$. To recover time in this setup, one can ---  following  Page and Wootters \cite{paw} ---
partition the Universe into a clock $C$ and the rest $R$, which are governed by Hamiltonians 
$\;\hat{H}_C={\rm Tr}_R\hat{H}\;$ and $\;\hat{H}_R={\rm Tr}_C\hat{H}$ respectively. 
       The conjugate to the clock Hamiltonian $\hat{P}_C$,
as defined   by  $\;[\hat{H}_C,\hat{P_C}]=i\;$ (with $\;\hbar=1$), plays a key role.
     Suppose that  $\;\hat{P}_C\ket{p}=p\ket{p}\;$ and that  $p$ is a continuous
variable (at least for all practical purposes). 
Then  $p$ can serve as the `evolution parameter' whose  associated evolution
operator is $\;U_C=e^{i\hat{H}_Cp}_C\;$.

        In order for the parameter $p$ to serve as `time' for both $C$ and $R$, their respective states should be maximally entangled; that is,  
        \begin{equation}
        \ket{\Psi}\;=\;\sum_p\alpha_{p}\ket{p}\ket{\phi_p}\;,
        \end{equation}
        where 
        $\ket{\phi_p}$ represents the states of $R$ (in terms of a suitable basis) and $\alpha_{p}$ represents complex  coefficients.
 
        Given that the Hamiltonian for the entire Universe vanishes, we have $\;\hat{H}=\hat{H}_C\otimes\mathbbm{1}+\mathbbm{1}\otimes\hat{H}_R+\hat{H}_I\;$, where $\hat{H}_I$ governs any interaction effects between $C$ and $R$. If $\hat{H}_I$ can be considered negligible, then $\;\hat{H}\approx\hat{H}_C\otimes\mathbbm{1}+\mathbbm{1}\otimes\hat{H}_R=0\;$, which leads to the relation
        \begin{equation}
         \hat{H}_C\;\approx\;-\hat{H}_R\;.
        \end{equation}
        This quasi-equality  makes it clear that  states of $C$ are correlated with those of $R$ in such a way that $p$ describes the evolution  of both.
        
    \section{Objections to the interacting clock solution}
    \label{b}
        \subsection{The clock ambiguity}
            The clock ambiguity has been  one of the main  criticisms
against the Page--Wootters method  \cite{amb}. Of particular interest to us
is not only the ambiguity itself but a recently proposed resolution that   is based  on using an isolated clock system \cite{ved}. Because of our emphasis on the interactions between the clock and the rest of the Universe, we would
like to propose a resolution that  does not depend on isolating the clock.
 
The clock ambiguity is just as the name suggests:  There are a multitude of 
ways, practically an infinite number,  to partition the Universe into a clock $C$ and the rest $R$.
It is then  safe to say that that different partitions  would generally  lead to different time parameters, as each choice of clock system 
will have its own unique  succession of states. This is simply not an acceptable 
state of affairs for an emergent notion of time. 

The choice of fully isolated subsystems does appear to circumvent this difficulty
because, as made clear in \cite{ved}, any such partition 
is related to all others
via a unitary transformation.  However, as shown in  the main text of
the current treatment, an interacting clock is an essential ingredient 
for the Page--Wootters framework to make sense. 
But, on the other hand, treating the isolated clock as  a limiting case of a more general situation, one  might may  be tempted to argue that the two views are
consistent provided that  the interactions are small enough. 
The problem with this argument  is that the effects of the interactions 
would  accumulate  and would have to  be taken into account eventually;  see Section~\ref{tt} and also \cite{us1,us2}.  However, the problem with
our counterargument is that the clock ambiguity must once again be confronted.
            
Yet,  in the framework of our discussion, any ambiguity in the choice of clock
is really besides the point.  This is because  there is no ambiguity in the arrow
of time that emerges by way of `recording' the interactions, and  this arrow naturally aligns with that of thermodynamics, 
which certainly has no ambiguity in its meaning of time. We would then propose
that the key to resolving the clock ambiguity is to make sure that one's
choice of  clock
is {\em not} isolated. 
            
        \subsection{Size discrepancies and  the large clock resolution}

Let us first note that one can expect, on generic grounds, that
 $\;\dim{C}\sim\dim{R}\;$ simply because of the condition of maximal entanglement. The same reasoning would imply that the two subsystems must basically agree 
on most measurable properties (such as the magnitude of each one's energy),
as maximal entanglement literally means that complete knowledge of one system
provides one with complete knowledge of the other. But let us suppose that
the dimensionalities do indeed differ. We can immediately discount a scenario like $\;\dim{C}\ll \dim{R}\;$ because any one configuration of the clock would 
then  correspond to a multitude of different  $R$ states, thus inhibiting the evolution of $R$. But what about the opposite situation? More importantly, could
one use the notion of a large clock to eliminate the need for interactions?

To address the second query, it is useful to recall from Section~\ref{tt} that isolated clocks can only provide a sense of time for a two-system Universe.
The basic point is that, given a clock $C$, 
the remainder of the Universe $R$ and  any parametrically small subsystem $R'$ of $R$,  
then $R'$ cannot have sufficient `knowledge' of the entanglement between the clock and its complement. On the other hand, if $R'$ is of the  same order in 
size  as  $R$, then there are still only two (or perhaps three) subsystems  
that are able to  experience time. 

 It might appear that this problem can be avoided by taking $C$ to be very large, $\;\dim{C}\gg\dim{R}\;$,
so that any other subsystem in the Universe could be as entangled as much
with the clock system as it could ever be. But this would not work because
each subsystem would then  `see'  a different sequence of clock configurations and,
therefore, would have its own  distinct  notion of time. For isolated subsystems this would indeed be the case --- but not if the interactions between the subsystems  are restored to provide
them with a universally agreed upon thermodynamic arrow of time. And so, although 
the case of  $\;\dim{C}\gg\dim{R}\;$ cannot be ruled out, it has no obvious selling points
and seems rather unnatural.

            \subsection{The multiverse and parallel universes}
                
We have, by intention, been restricting considerations to 
the case for which the  Universe is in isolation. If one is invested in models
with a  large number  of parallel universes and/or  the now popular
multiverse framework, then these external `verses' would have
to be  precluded from 
influencing the Universe in question. Alternatively, should one or more
of these external verses be shown  to influence the Universe of interest,
then the boundary of the system could be extended to include the influencing systems. Only an infinite number of such influencers would then produce any 
significant  obstruction to our discussion.

            \subsection{The Block Universe}
            
The so-called Block Universe is not too 
far removed from the timeless models of the Universe which were considered,
and then mostly  dismissed, in  Section~\ref{dis}. Nevertheless, there is enough of
a distinction to warrant a separate comment.

There are actually many variants of  Block Universe
(see, {\em e.g.},  \cite{daiton}, for a summary), 
but any of these describe the same basic picture:
A deterministic reality in which all `moments of time' can be viewed as spacelike slices, with each one 
stacked on another to  form a never-changing
 four-dimensional `block' of spacetime.  There is no possibility of distinguishing between past, present and future, as all such slices are meant to
be equal in status. 
Consequently, any experience of time  or any  description of a transitory `now' moment should be  regarded as an illusion.~\footnote{The same idea is also described by McTaggart's B-series of time \cite{mcT}.} As the concept of  time
  becomes trivialized, there is indeed a sense of timelessness for these models.
This timelessness  is, however, different from that of our frozen 
Universe because
the former cannot incorporate causality ---  its inclusion would inevitably require one to treat change as a real physical process. This point is elaborated on in 
Section~4.


\begin{thebibliography}{1}

    %1
    \bibitem{time1}
    S.~Hawking and R.~Penrose, {\em The nature of space and time} 
(Princeton University Press, 2010).

    %2    
    \bibitem{time2}
    J.~Barbour, ``The nature of time,'' [arXiv:0903.3489 [gr-qc]].

    %3    
    \bibitem{time3}
    H.~Price, ``Cosmology, time's arrow, and that old double standard,'' [arXiv:9310022 [gr-qc]].
    
    \bibitem{time4}
    C.~Rovelli, {\em The Order of Time} (Penguin Books, 2018).

    \bibitem{norton}
    J.~D.~Norton, ``Time really passes,'' Humana. Mente: Journ. of Phil. St. 
{\bf 13} (2010).
    


    \bibitem{daiton}
    B.~Dainton, {\em Time and space} (Routledge, 2016).
    %4  
    \bibitem{popper}
    K.~R.~Popper, {\em The world of Parmenides}, (Taylor \& Francis,  2012).
   
    \bibitem{phil}
    M.~Nelson, ``Existence,'' in {The Stanford Encyclopedia of Philosophy},
 Winter 2016 Ed., Edward N. Zalta, ed., \\ 
https://plato.stanford.edu/archives/win2016/entries/existence/.
   
    %5

    
    \bibitem{sider}
    T.~Sider, {\em Four-dimensionalism: An ontology of persistence and time} 
(Oxford University Press, 2001).
    
    %6   
    \bibitem{pearle}
     J.~Pearl, {\em Causality} (Cambridge University Press, 2009).
     
    %7 
    \bibitem{white}
    A.~N.~Whitehead and D.~W.~Sherburne, {\em Process and reality} 
(Macmillan, New York, 1957).
        
    %8
    \bibitem{prig}
    I.~Prigogine, {\em From being to becoming: Time and complexity in physical systems} (W.H. Freeman, San Francisco, 1980).    
    
    %9
    \bibitem{QMcaus}
    G.~M.~D’Ariano, ``Causality re-established,'' 
Phil. Trans. R. Soc. A {\bf 376}, 20170313 (2018).
    
    \bibitem{albert}
    D.~Z.~Albert, {\em Time and chance} (Harvard University Press, 2003).    

    \bibitem{ellis}
    G.~F.~R.~Ellis, ``On the flow of time,'' [arXiv:0812.0240 [gr-qc]].
    
    %10
    \bibitem{ar1}
    S.~W.~Hawking, ``The Direction of Time,'' New Scientist {\bf 115}, 
 46 (1987).
    
    %11
    \bibitem{ar2}
    J.~B.~Hartle, ``The physics of ’now’,'' Amer. Jour. of Phys. {\bf 73}, 
101 (2005).

    %12
    \bibitem{lan}
    R.~Landauer, 
``Irreversibility and heat generation in the computing process,'' IBM journal of research and development {\bf 5}, 183 (1961).
    
    %13
    \bibitem{wdw}
      B.~S.~DeWitt, ``Quantum theory of gravity. I. The canonical theory,'' Phys.\ Rev.\ {\bf160}, 1113 (1967). 
%doi:10.1103/PhysRev.160.1113.
    
    %14      
    \bibitem{paw}
      D.~N.~Page and W.~K.~Wootters, ``Evolution without evolution: Dynamics described by stationary observables,'' Phys.\ Rev.\ D {\bf27},  2885 (1983). 
%doi:10.1103/PhysRevD.27.2885.
    
    %15      
    \bibitem{us1}
      K.~L.~H.~Bryan and A.~J.~M.~Medved, ``Realistic clocks for a Universe without time,'' Found.\ Phys.\  {\bf 48},  48 (2018)
% doi:10.1007/s10701-017-0128-x 
[arXiv:1706.02531 [quant-ph]].
    
    %16
    \bibitem{us2} 
      K.~L.~H.~Bryan and A.~J.~M.~Medved, ``Requiem for an ideal clock,'' [arXiv:1803.02045 [quant-ph]].
    
    %17
    \bibitem{amb}
    A.~Albrecht and A.~Iglesias, ``The Clock ambiguity and the emergence of physical laws,'' Phys.\ Rev.\ D {\bf 77}, 063506 (2008)
 %   doi:10.1103/PhysRevD.77.063506 
[arXiv:0708.2743 [hep-th]].
    
    %18
    \bibitem{newt}
    S.~Hawking, {\em A Brief History of Time} (Bantam, 1988).
    
    %19
    \bibitem{gr}
    R.~d'Inverno, {\em Introducing Einstein's Relatvity} 
(Oxford University Press, 1992).
    
    %20
    \bibitem{str}
    J.~Polchinski, {\em String theory. Vol. 1: An introduction to the bosonic string, Vol. 2: Superstring theory and beyond} (Cambridge University Press,  
1998).
    
    %21
    \bibitem{ww}
    S.~Weinberg and E.~Witten, ``Limits on massless particles,'' Phys.\ Lett.\ B {\bf 96}, 59  (1980).
    
    %22    
    \bibitem{wit}
      E.~Witten, ``Symmetry and Emergence,'' Nature Phys.\  {\bf 14}, 116 
(2018).
% doi:10.1038/nphys4348.
    
    %23  
    \bibitem{wit1}
    G.~’t Hooft, ``Symmetry Breaking Through Bell-Jackiw Anomalies,'' Phys.\ Rev.\ Lett. {\bf37},  8 (1976).
    
    %24
    \bibitem{wit2}
    T.~Banks and L.~J.~Dixon, ``Constraints on string vacua with spacetime supersymmetry,'' Nucl.\ Phys.\ B {\bf 307}, 93  (1988).
    
    %25
    \bibitem{st}
      J.~Polchinski, ``M theory: Uncertainty and unification,'' [arXiv:0209105[hep-th]].
    
    %26
    \bibitem{beken}
    J.~D.~Bekenstein, ``Nonexistence of baryon number for static blackholes,'' Phys. Rev. D {\bf 5}, 1239 (1972).
   
    %27  
    \bibitem{page}
    D.~N.~Page, ``Particle emission rates from a black hole. I: Massless particles from an uncharged, nonrotating hole,'' Phys. Rev. D {\bf 13}, 198 (1976).
    
    %28
    \bibitem{kuch}
     K.~V.~Kuchar, ``Time and interpretations of quantum gravity,''
in {\em Proc. 4th Canadian Conference on General Relativity and Relativistic Astrophysics}, G.~Kunstatter, D.~E.~Vincent and J.~G.~Williams, eds.
(1992). 
%doi:10.1142/S0218271811019347.
    
    %29
    \bibitem{cc}
     C.~E.~Dolby, ``The Conditional probability interpretation of the Hamiltonian constraint,'' [arxiv:0406034 [gr-qc]].
   
    %30
    \bibitem{ll}
     V.~Giovannetti, S.~Lloyd and L.~Maccone, ``Quantum Time,'' Phys.\ Rev.\ D {\bf 92}   045033 (2015)
% doi:10.1103/PhysRevD.92.045033 
[arXiv:1504.04215 [quant-ph]].
    
    %31
    \bibitem{vil}
    A~.Vilenkin, ``Quantum cosmology and the initial state of the universe,''  Phys. Rev. D {\bf 37}, 888 (1988).
     
     %32
     \bibitem{ved2}
      S.~Stupar and V.~Vedral, ``Was inflation necessary for the existence of time?,''
  [arXiv:1710.04260 [quant-ph]].
  
%     \bibitem{gold}
%     Gold, Th. (Ed.). (1967). The nature of time. Ithaca: Cornell University Press
  
    %33
    \bibitem{mcT}
    J.~E.~McTaggart, ``The unreality of time, ''  Mind {\bf 17},  457 (1908).
    
    \bibitem{hume}
    D.~Hume, ``An enquiry concerning human understanding,'' in {\em Seven Masterpieces of Philosophy}, 191 (Routledge, 2016).
    
    %34
    \bibitem{smolin}
     L.~Smolin, ``Temporal relationalism,'' arXiv:1805.12468 [physics.hist-ph].

    %35
    \bibitem{carrol}
  S.~M.~Carroll, ``What if Time Really Exists?,'' 
  [arXiv:0811.3772 [gr-qc]] (2008). 
     
     \bibitem{bouton}
     C.~Bouton, ``Is the Future already Present? The Special Theory of Relativity and the Block Universe View,'' in  {\em 
Time of Nature and the Nature of Time},  
89  (Springer, 2017).
     
     \bibitem{erasmus}
     J.~Erasmus, ``Can Cosmology Justify Belief in an Eternal Universe?,'' in
 {\em The Kalam Cosmological Argument: A Reassessment},  129 
(Springer, 2018).
     
     \bibitem{frisch}
     M.~Frisch, {\em Causal reasoning in physics}
(Cambridge University Press, 2014).
     
    %36
     \bibitem{jac}
    T.~Jacobson, ``Thermodynamics of spacetime: the Einstein equation of state,'' Phys.\ Rev.\ Lett.\ {\bf 75}, 1260 (1995). 
%doi:10.1103/PhysRevLett.75.1260.
    
\bibitem{Ryu} 
S.~Ryu and T.~Takayanagi,
  ``Holographic derivation of entanglement entropy from AdS/CFT,''
  Phys.\ Rev.\ Lett.\  {\bf 96}, 181602 (2006)
  %doi:10.1103/PhysRevLett.96.181602
  [hep-th/0603001].

    %37    
    \bibitem{dualities}
      O.~Aharony, S.~S.~Gubser, J.~M.~Maldacena, H.~Ooguri and Y.~Oz, ``Large N field theories, string theory and gravity,'' Phys.\ Rept.\  {\bf 323}, 183 (2000) 
%doi:10.1016/S0370-1573(99)00083-6
    [hep-th/9905111].
    
    %38    
    \bibitem{ads}
      N.~Engelhardt and S.~Fischetti, ``Losing the IR: a Holographic Framework for Area Theorems,''
    [arXiv:1805.08891 [hep-th]].
    
    %39   
  \bibitem{ved}
      C.~Marletto and V.~Vedral, ``Evolution without evolution and without ambiguities,'' Phys.\ Rev.\ D {\bf 95}  043510 (2017) 
%doi:10.1103/PhysRevD.95.043510 [arXiv:1610.04773 
[quant-ph]].

      
      
      %
%  \bibitem{wheel}
%   J.~A.~Wheeler, ``Information, physics, quantum: The search for links,'' Complexity, entropy, and the physics of information 8 (1990).
%   
%  \bibitem{zurek}
%       W.~H.~Zurek, ``Quantum Darwinism and Envariance,'' [arXiv:0308163 [quant-ph]]
% %      
%     \bibitem{pen}
%      R.~Penrose, ``The Emperor's New Mind,'' Oxford University Press (1989)  
%     \bibitem{th}
%       G.~t.~Hooft, ``Time, the arrow of time, and Quantum Mechanics,'' [arXiv:1804.01383 [quant-ph]]
% %  
%     \bibitem{na}
%       S.~Das and G.~Kunstatter, ``The central role of symmetry in physics,'' Journal of Applied and Fundamental Sciences (Assam Don Bosco University, India), Vol.2(2), 69-77 (2016) [arXiv:1609.02038 [gr-qc]]
% %  
%     \bibitem{ads}
%     M.~Van Raamsdonk, ``Lectures on Gravity and Entanglement,'' [arXiv:1609.00026 [hep-th]] 
%
%     \bibitem{ds}
%      A.~Strominger, ``The dS / CFT correspondence,'' JHEP {\bf 0110}, 034 (2001) doi:10.1088/1126-6708/2001/10/034 [arXiv:0106113[hep-th]]
%
    %
    %
      
\end{thebibliography}
\end{document}